# Thermally activated diffusion and lattice relaxation in (Si)GeSn materials

Nils von den Driesch[1,2], Stephan Wirths[2], Rene Troitsch[2], Gregor Mussler[2], Uwe Breuer[3], Oussama Moutanabbir[4], Detlev Grützmacher[1,2] and Dan Buca[2]

[1]JARA-Institut Green IT, RWTH Aachen, Sommerfeldstraße 14, 52074 Germany

[2]Peter Grünberg Institute (PGI 9) and JARA-FIT, Forschungszentrum Jülich, 52425 Jülich, Germany

[3]Zentralinstitut für Engineering, Elektronik und Analytik (ZEA 3), Forschungszentrum Jülich, 52425 Jülich, Germany

[4]Department of Engineering Physics, École Polytechnique de Montréal, Montréal, C.P. 6079, Succ. Centre-Ville, Montréal, Québec, Canada H3C 3A7

**Abstract**

Germanium-Tin (GeSn) alloys have emerged as a promising material for future optoelectronics, energy harvesting and nanoelectronics owing to their direct bandgap and compatibility with existing Si-based electronics. Yet, their metastability poses significant challenges calling for in-depth investigations of their thermal behavior. With this perspective, this work addresses the interdiffusion processes throughout thermal annealing of pseudomorphic GeSn binary and SiGeSn ternary alloys. In both systems, the initially pseudomorphic layers are relaxed upon annealing exclusively via thermally induced diffusional mass transfer of Sn. Systematic post-growth annealing experiments reveal enhanced Sn and Si diffusion regimes that manifest at temperatures below 600°C. The amplified low-temperature diffusion and the observation of only subtle differences between binary and ternary hint at the unique metastability of the Si-Ge-Sn material system as the most important driving force for phase separation.

**Introduction**

The Si-Ge-Sn semiconductor system has sparked a great deal of interest among scientists because of its unique properties within group IV materials. The demonstrated fundamental direct bandgap of this truly silicon-compatible material can pave the way for numerous new applications in the field of opto- and nanoelectronics [1]. Research concerning GeSn-based light emitters was especially spurred by the demonstration of optically pumped lasing in bulk [2,3] and heterostructure layers [4,5] and may one day lead to the convergence of electronics and photonics circuitry [6,7]. Nevertheless, plenty of obstacles in this inherently metastable material system must be overcome to lay the groundwork for real applications.

One of the main challenges posed by (Si)GeSn alloys is maintaining their structural integrity during thermal treatment in order to preserve their intrinsic material properties. For example, thermal budget must be specifically kept low enough to avoid Sn diffusion out of the material and associated segregation [8]. This physical process yields a phase separation with the GeSn equilibrium phase at a Sn content well below the critical value for a direct bandgap, rendering it unsuitable for light emitting devices. Moreover, thermally activated interdiffusion processes can smear out the interfaces and hence drastically alter the basic heterostructure properties, such as band offsets or confinement potential.

Despite this critical importance, detailed studies of thermal behavior of metastable (Si)GeSn materials are still conspicuously missing in literature. In contrast, an extensive body of knowledge is available on the canonical binary group IV system Si-Ge. In fact, it has been established that – depending on the initial layer strain and dislocation density in the material – both diffusion and strain relaxation via dislocation formation take place upon annealing [9,10] and the former being more relevant in coherent layers. Similarly, the mechanism governing thermal relaxation behavior in (Si)GeSn alloys is indistinct, counterintuitive, and not fully understood yet, encompassing Sn diffusion, phase segregation, and defect nucleation [11–15]. The observed rather low thermal stability of GeSn, compared to SiGe alloys, however, originates expectedly from the low solid solubility of Sn in Ge, making even unstrained layers thermodynamically metastable [16]. This inherent material property is the driving force for low-temperature Sn diffusion, for the formation of Sn-rich precipitates on the surface and nano-crystals inside the material upon annealing, since a phase separation into elemental Ge and β-Sn is thermodynamically favored [11,17,18]. However, the role of strain in altering the thermodynamic properties of GeSn layers is yet to be determined. There is even less certainty for the case of ternary SiGeSn alloys. Fundamentally, phase separation in ternary alloys should differ from the binary case, since a larger possible number of intermediate phases can be formed [19]. The solubility of Sn in a SiGe matrix is even smaller compared to a Ge one [16], which may point to less stable ternaries in that case. On the other hand, however, the enhanced mixing entropy in ternary alloys was proposed to explain a predicted stability enhancement [20].

In this work, we investigate the diffusion and lattice relaxation of both pseudomorphic GeSn binary and SiGeSn ternary alloys upon thermal annealing, to identify the behavior of each one of those systems and reveal possible differences between them. Epitaxial growth of thin films were conducted at temperatures below 400°C in a commercial AIXTRON Tricent reduced-pressure chemical vapor deposition (CVD) reactor employing a reactive gas source epitaxial process using the precursors digermane ($Ge_2H_6$), disilane ($Si_2H_6$) and tin tetrachloride ($SnCl_4$) [21]. Alloys were grown on Ge buffers to minimize the large lattice mismatch to the underlying 200 mm Si(001) wafers. Two pairs of GeSn and SiGeSn samples with matching Sn concentrations of roughly 6 at.% and 9 at.%, respectively, were prepared to allow a direct comparison between binary and ternary layers. The thickness of all layers was beneath 50 nm, chosen well below the critical thickness for plastic strain relaxation, to suppress the impact of defects on the diffusion experiments. The pair of GeSn and SiGeSn layers, which was intended for determining the diffusion constants, was capped by additional 300 nm and 150 nm Ge, respectively. The cap layers were grown at the very same temperature as the binary and ternary to rule out an influence of the sample surface on the diffusion process. For the experiments, samples were annealed in an industrial Mattson Helios rapid thermal processing (RTP) system in a temperature range between 350-850°C under nitrogen atmosphere for different periods of time between 3 and 30 minutes.

**Results**

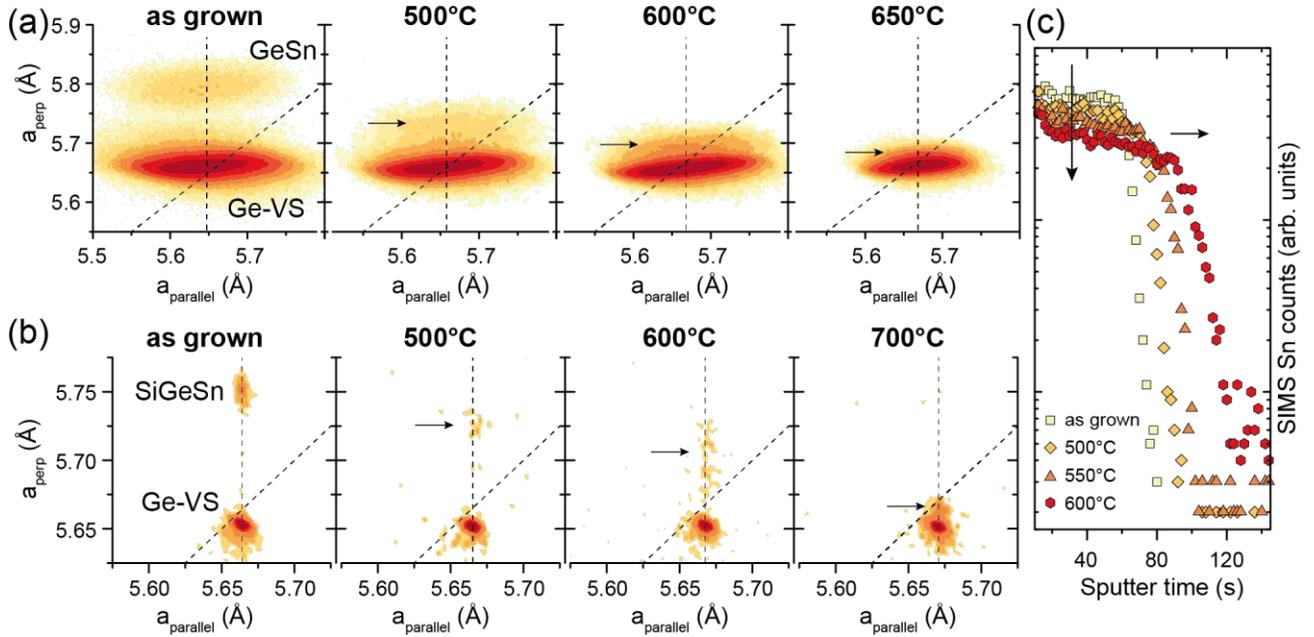

*Figure 1. XRD Reciprocal space maps (RSM) of coherently grown a) Ge$_{0.914}$Sn$_{0.086}$ and b) Si$_{0.10}$Ge$_{0.81}$Sn$_{0.09}$ samples, as well as c) SIMS spectra of the former, annealed at different temperatures.*

To elucidate the thermal behavior of (Si)GeSn semiconductors, systematic annealing experiments were performed on Ge$_{0.914}$Sn$_{0.086}$ and Si$_{0.10}$Ge$_{0.81}$Sn$_{0.09}$ samples, pseudomorphically grown on top of Ge buffers. The buffer layer exhibits a very low threading dislocation density (<1E7 cm$^{-2}$) [22], which defines the upper limit of threading arms gliding through the pseudomorphic GeSn layer. Reciprocal space maps (RSM) of the former, obtained from X-ray diffraction (XRD) experiments around the asymmetric (224) reflection, are shown for different annealing temperatures in **Figure 1**a. Coherent growth is proven by the matching in-plane lattice constants $a_{parallel}$ of GeSn layer and

Ge virtual substrate (Ge-VS). Annealing of the samples for thirty minutes at temperatures below 450°C does not have a distinct effect on the layer strain, which is consistent with thermal stability data on alloys with similar composition [11]. Subsequent to thermal annealing at 500°C for 10 minutes, however, the GeSn peak moves towards smaller out-of-plane lattice constants $a_{perp}$, closer to the Ge-VS lattice. Simultaneously, the in-plane lattice constant remains fixed at the Ge-VS value. Such lattice parameter evolution indicates that, although Sn diffuses out of the layer, the GeSn layer always remains coherently strained on top of the buffer.

By examining the XRD data recorded on the pseudomorphic $Si_{0.10}Ge_{0.81}Sn_{0.09}$ layers (Fig. 1b), the ternary was found to behave qualitatively similarly to the binary case (Fig. 1a). Also, the gradual diffusion of Sn upon annealing causes a shift of the SiGeSn XRD signal towards the buffer, yet remaining a coherently strained layer. In both samples, annealing above 700°C decreases the mean Sn content to 1 at.%. In such mixture, a thermodynamically stable solid solution is formed, which triggers strongly enhanced Sn diffusion, as we will discuss later on.

Diffusion of Sn in samples can be easily made visible from secondary ion mass spectroscopy (SIMS) profiles, which are displayed in Fig. 1c. These measurements unravel the kinetics of Sn diffusion throughout thermal annealing for 10 minutes at different temperatures. Besides the expected spread of Sn in larger depths, some segregation of Sn is also observed on the surface above 600°C. Interestingly, the Sn composition profiles itself remain rather rectangular during annealing. Sn atoms diffuse roughly 40 nm into the Ge buffer when annealed at 600°C for 10 minutes, yet, it only reduces the Sn concentration of the layer and smears the GeSn/Ge interface. No Gaussian shape of the composition profiles, as expected from a composition-independent diffusion coefficient, can be seen in our experiments.

In general, the observed gradual diffusion of Sn in pseudomorphic layers stands in distinct contrast to the recorded behavior in partially relaxed material. In the latter, thermal annealing leads to an abrupt transition to the two equilibrium phases [11,23–25]. This is linked to a preferential diffusion of Sn atoms along dislocations [26], enhancing the clustering of Sn atoms that ultimately leads to the formation of β-Sn crystallites and precipitates on the surface and in bulk. On the other hand, as seen above, Sn diffusion in pseudomorphic layers is a gradual process, depending strongly on temperature and annealing times.

For diffusion experiments, 30 nm thin $Ge_{0.94}Sn_{0.06}$ and $Si_{0.040}Ge_{0.895}Sn_{0.065}$ samples were grown coherently on top of Ge-buffered Si(001) wafers. To investigate diffusion into the surrounding Ge matrix and suppress surface segregation, which will influence the former as for the samples in figure 1, both GeSn and SiGeSn layers were capped in-situ with 300 nm and 150 nm thick Ge layers, respectively.

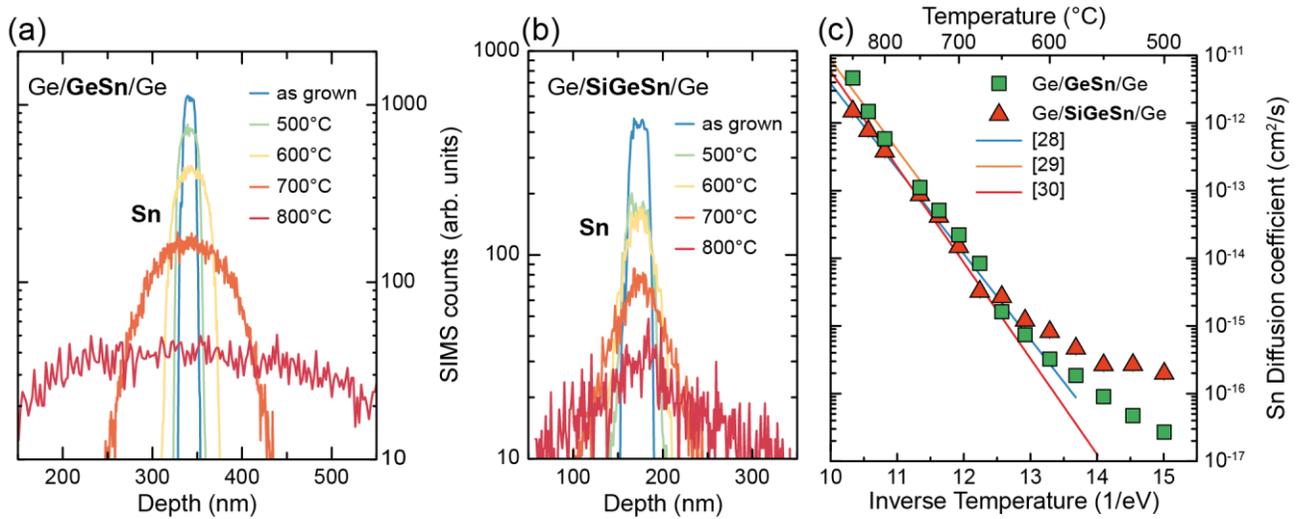

*Figure 2. Sn diffusion in pseudomorphically grown a) $Ge_{0.94}Sn_{0.06}$ and b) $Si_{0.040}Ge_{0.895}Sn_{0.065}$, sandwiched between two layers of Ge, respectively. c) Comparison of Sn diffusion coefficients to literature data.*

**Figure 2**a depicts the diffusion of Sn atoms in the GeSn structure after annealing for different periods between 3 and 10 minutes at temperatures between 500°C and 800°C, as no distinct effect was observed at lower temperatures. In the as-grown sample, Sn atoms exhibit a rather rectangular

distribution, hinting to a good GeSn layer quality with sharp interfaces. Up to 600°C annealing, the Sn profiles look flat and with steeper sides than one would expect for a constant diffusion coefficient. Annealing above 600°C transforms the profiles into a more Gaussian-like shape. Thereupon, Sn content in the layer is strongly reduced – more than a factor of two after annealing at 600°C for 10 minutes – compared to the as-grown case. A qualitatively similar behavior can be seen for Sn diffusion in a SiGeSn ternary, as shown in Figure 2b, indicating an increased Sn interdiffusion into the surrounding Ge. In comparison to the binary, however, Sn diffusion in SiGeSn sets in relatively easier, as the Sn peak concentration decreases more rapidly at lower temperatures.

The distinct diffusion behavior of Sn in the binary and ternary (Si)GeSn materials are summarized in Figure 2c. In this figure, Sn diffusion coefficients, determined by fitting appropriate solutions of the one-dimensional diffusion equation [27], are plotted against $(k_BT)^{-1}$ in a large temperature range between 500-850°C. Low- and high-temperature regimes can clearly be distinguished from varying slopes in the Arrhenius plot. Above ~650°C, diffusion of Sn atoms from GeSn, as well as SiGeSn, is similar to reports on diffusion of implanted Sn into Ge wafers [28–30], as indicated by lines in Figure 2c. The temperature-dependent behavior of the Sn diffusion coefficient in this region $D(T)$ can be described by an Arrhenius behavior

$$D(T) = D_0 \times \exp\left(-\frac{E_A}{k_BT}\right),$$

with a constant pre-exponential factor $D_0$, activation energy $E_A$ and Boltzmann constant $k_B$. An overview on the obtained parameters can be found in Table I. The activation energies of 3.10 ± 0.07 eV (GeSn) and 2.87 ± 0.03 eV (SiGeSn) are still in line with literature values, where a spread between 3.26 eV and 2.90 eV has been reported for ion-implanted Sn impurities in Ge [28–30]. A slightly smaller activation energy in the case of SiGeSn may indicate a larger number of vacancies in that material, as we will discuss in the upcoming section. The overall diffusion coefficient above 650°C, however, is always slightly smaller than in GeSn. Likely the presence and diffusion of Si atoms, which also occurs by a vacancy-mediated mechanism [31], impacts diffusion of the second species and leads to diverging activation energies.

In the low temperature regime below ~650°C an interesting change in diffusion behavior becomes apparent. The temperature dependence is less pronounced and cannot be described by the same activation energies anymore. Reduced activation energies of $1.81 \pm 0.09$ eV (GeSn) and $1.13 \pm 0.10$ eV (SiGeSn) indicate strongly enhanced diffusion in that regime. Such enhancement may be predominantly driven by different physical origins, such as the concentration gradient or strain. Importantly, those can manifest in the thermal behavior only due to the absence of dislocations, which would otherwise overshadow their effects. The individual impact of those driving forces will be reconsidered in the *Discussion* section.

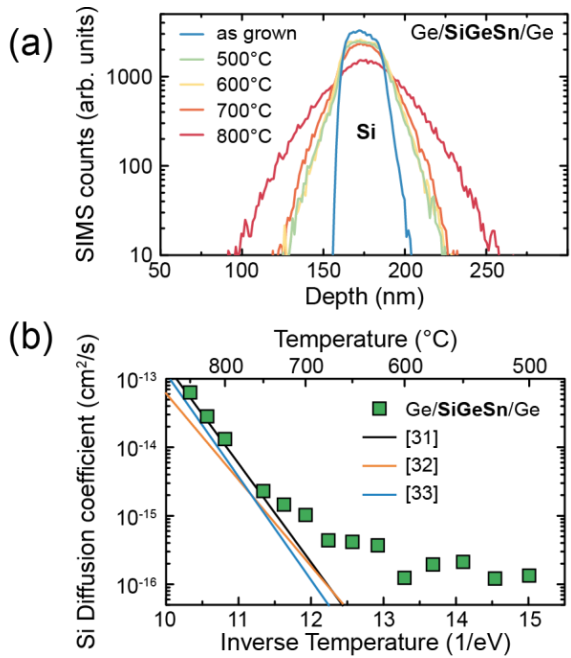

***Figure 3.*** *a) Si diffusion from $Si_{0.040}Ge_{0.895}Sn_{0.065}$ and b) comparison of the derived diffusion coefficients to literature data.*

Diffusion of Si atoms, depicted in **Figure 3**a, shows a similar behavior compared to Sn diffusion. Again, it consists of two different regimes above and below ~700°C, as demonstrated by the diffusion coefficients in Fig. 3b. This observation is rather surprising, since previous works concerning Si diffusion in Ge always found a behavior well described by a single activation energy down to 650°C [32]. Indeed, the determined activation energy of $3.27 \pm 0.03$ eV in the high temperature regime very well matches literature values of vacancy-mediated Si diffusion in Ge, which spread between 2.90 eV and 3.47 eV [31–33]. The larger activation energy compared to Sn diffusion is typically attributed to a size effect. The smaller Si atom size results in a reduced attractive interaction with vacancies, which determine diffusion behavior of group IV elements in Ge lattices[31]. In the low-temperature regime, the activation energy drops down to 0.50 eV.

**Table I**: Overview of determined activation energies $E_A$ and pre-exponential factors $D_0$ for diffusion of Sn and Si from GeSn and SiGeSn layers in the high and low temperature regimes.

| Regime | Material stack | Sn | | Si | |
|---|---|---|---|---|---|
| | | $E_A$ (eV) | $D_0$ (cm$^2$/s) | $E_A$ (eV) | $D_0$ (cm$^2$/s) |
| High T | Ge/GeSn/Ge | $3.10 \pm 0.07$ | $264 \pm 220$ | - | - |
| | Ge/SiGeSn/Ge | $2.87 \pm 0.03$ | $12 \pm 5$ | $3.27 \pm 0.03$ | $31 \pm 9$ |
| Low T | Ge/GeSn/Ge | $1.81 \pm 0.09$ | - | - | - |
| | Ge/SiGeSn/Ge | $1.13 \pm 0.11$ | - | $0.50 \pm 0.10$ | - |

**Discussion**

An important observation in our experiments is that only isolated Sn diffusion occurs during thermal annealing in both binary and ternary alloys. In literature, also strain relaxation upon thermal annealing via defect formation was observed for both initially relaxed GeSn material [34], as well as for coherently strained GeSn films on rather thin Ge buffer layers [14,35]. We attribute the different behavior upon annealing to differences in initial crystallinity and buffer quality. In case of low initial threading dislocation densities (TDD), as it is the case for our coherently grown material, the formation of isolated defects is energetically not favored. Thus, thermal strain relaxation occurs rather elastically via diffusion of Sn atoms, as demonstrated by RSM of the annealed samples in Figure 1. Grown-in dislocations, on the other hand, are present in partly relaxed GeSn or coherently grown material on top of Ge buffers with larger TDDs. In those, multiplication and gliding of pre-existing dislocations are kinetically enabled during thermal annealing. This mechanism generates an array of strain-relieving misfit segments, which is well known from relaxation via dislocations in SiGe alloys [36]. Furthermore, the generation of dislocations also changes the observed phase separation behavior of GeSn alloys, which is part of an ongoing investigation that will be reported elsewhere.

Generally, diffusion in a multi-component system is driven by a minimization of the Gibbs free energy. Therefore, concentration gradients between the solute and the surrounding matrix will strongly impact interdiffusion, which was shown for example in the Si-Ge system [37,38]. On the other hand, also strain inside the material may enhance diffusion towards an equilibrium state via mass transfer, as indicated in literature [10,39]. Additionally, the thermodynamic instability of the (Si-)Ge-Sn material system may also enhance Sn diffusion out of the binary, as it moves the system closer to its thermodynamic equilibrium composition (below ~1 at.% Sn) [40].

The emergence of two clearly separated temperature regimes of the Sn diffusion coefficient in our work is in stark contrast to previous studies reported in literature. We attribute the presence of the low-temperature regime to the metastability of the Si-Ge-Sn material system, which is unique among the group IV alloys. Above the transition temperature of roughly 650°C (see Fig. 2c), the remaining (mean) Sn concentrations in the alloys approach the Sn solid solubility limit of 1 at.% (c.f. Fig. 1a and b). Accordingly, the extracted activation energies for Sn diffusion match those previously reported for impurity diffusion in a Ge host lattice [28–30].

In the low-temperature regime, however, different physical effect are possible origins of the enhanced diffusion. Material strain, for example, is known to reduce the formation energy of point defects, which mediate diffusion processes. In case of compressively strained SiGe, the formation energy of vacancies is decreased, leading to enhanced Ge interdiffusion [39]. Therefore, compressive strain will likely also have an impact on vacancy-mediated Sn diffusion in the low-temperature regimes, but cannot be the sole explanation. Since our investigated $Ge_{0.94}Sn_{0.06}$ and $Si_{0.040}Ge_{0.895}Sn_{0.065}$ thin films share the same compressive strain values (-0.90 % and -0.89 %, respectively) the activation energy reduction should be similar. Instead, the reduction is much more pronounced for the ternary (2.87 eV to 1.13 eV) compared to the binary case (3.10 eV to 1.81 eV). Therefore, the metastability of Sn-based alloys, i.e. the gradient in chemical potential, is the main driving force, which triggers the diffusion enhancement in both binary and ternary. Since the solubility of Sn inside SiGe is even smaller than in elemental Ge – making SiGeSn ternaries even more metastable alloys [16] – the chemical potential gradient can also explain the observed deviations between binary and ternary, namely reduced activation energies for the latter case.

The drop of activation energy for Si diffusion in the low temperature regime, which has never been reported for diffusion in an elemental Ge matrix, may also be linked to the presence of Sn atoms. From atomistic-level analysis it is known that a repulsive interaction between Si and Sn atoms is present in ternary SiGeSn alloys, leading to deviations from a random distribution [41]. This additional effect may influence Si diffusion in the early stages, before a greater dilution of Si atoms is reached at higher temperatures.

Another factor, which role is not fully understood yet, is the expected presence of vacancies inside (Si)GeSn alloys. Those are inherently present due to the low temperature epitaxy of thin films and have recently been directly observed using positron annihilation in GeSn alloys [42]. Since vacancies can locally relax the strain surrounding the larger size Sn atoms, an attractive interaction between Sn atoms and vacancies is expected in GeSn alloys, however, the impact of Si has not yet been investigated. Therefore, little is known about differences in formation and migration energies of vacancies between GeSn and SiGeSn alloys, which thus cannot be ruled out as possible origin for the different diffusive behavior.

**Conclusions**

In this work, we investigated the effect of thermal annealing on pseudomorphic GeSn binary and SiGeSn ternary alloys. In the absence of dislocations, strain is relaxed only via thermally-activated interdiffusion of Sn atoms. The qualitatively same behavior for binaries and ternaries was found in systematic annealing experiments, revealing an additional strongly enhanced Sn diffusion regime at low temperatures (<650°C). Similar behavior, albeit less pronounced, was found for diffusion of Si atoms, pinpointing the unique metastability of the group IV Si-Ge-Sn material system as the most important driving force for interdiffusion. Our results underline the importance of carefully handling the thermal budget of (Si)GeSn layers to maintain their pristine material properties during (thermal) processing in order to obtain highly efficient group IV-based opto- and nanoelectronic devices.